# Measuring coronal magnetic fields with remote sensing observations of shock waves

A. Bemporad[1], R. Susino[1], F. Frassati[1], S. Fineschi[1]

[1]INAF, Turin Astrophysical Observatory, Pino Torinese (TO), Italy

**Abstract**

Recent works demonstrated that remote sensing observations of shock waves propagating into the corona and associated with major solar eruptions can be used to derive the strength of coronal magnetic fields met by the shock over a very large interval of heliocentric distances and latitudes. This opinion article will summarize most recent results obtained on this topic and will discuss the weaknesses and strengths of these techniques to open a constructive discussion with the scientific community.

## 1. Introduction

### 1.1. State of the art

Our limited knowledge of the magnetic fields structuring in the solar corona represents today the main hurdle in our understanding of its structure and dynamic. Over the last decades significant efforts have been dedicated to measure these fields, by approaching the problem on many different sides and in particular: i) by improving our theoretical understanding of the modification (via Zeeman and Hanle effects) induced by these fields on the polarization of coronal emission lines, ii) by developing new instrumentation to measure directly with spectro-polarimeters these modifications, iii) by improving the reliability of the extrapolated coronal fields starting from photospheric measurements, iv) by developing new techniques to analyse existing remote sensing data and infer properties of these fields, or by combining all these different approaches (e.g. Chifu et al. 2015).

In this paper we focus on the fourth method, discussing the advantages and disadvantages of techniques recently developed. Over the last few years it has been shown that coronagraphic white light (WL) observations of major Coronal Mass Ejections (CMEs) show the presence of hemispherical regions expanding ahead of the CME fronts (Figure 1) where faint increases in the WL intensity are detected (e.g. Vourlidas et a. 2003; Ontiveros & Vourlidas 2009). More recently similar features have also been detected in the early stages of CME developments observed with EUV imagers (e.g. Kozarev et al. 2011). Usually these slightly brighter regions are interpreted as the coronal plasma compressed by the transit of the shock wave driven by the super-alfvénic expansion of the CME, a region called "shock sheath". This interpretation has also been supported by MHD simulations (e.g. Manchester et al. 2008) and forward modelling of synthetic observations (Vourlidas et al. 2013) showing the formation of an arch shaped feature very similar to what is observed in WL coronagraphs. In this scenario, it has been recently demonstrated that measurements of coronal fields can be successfully derived from remote sensing observations of interplanetary shocks with at least two techniques, briefly described in the next Sections.

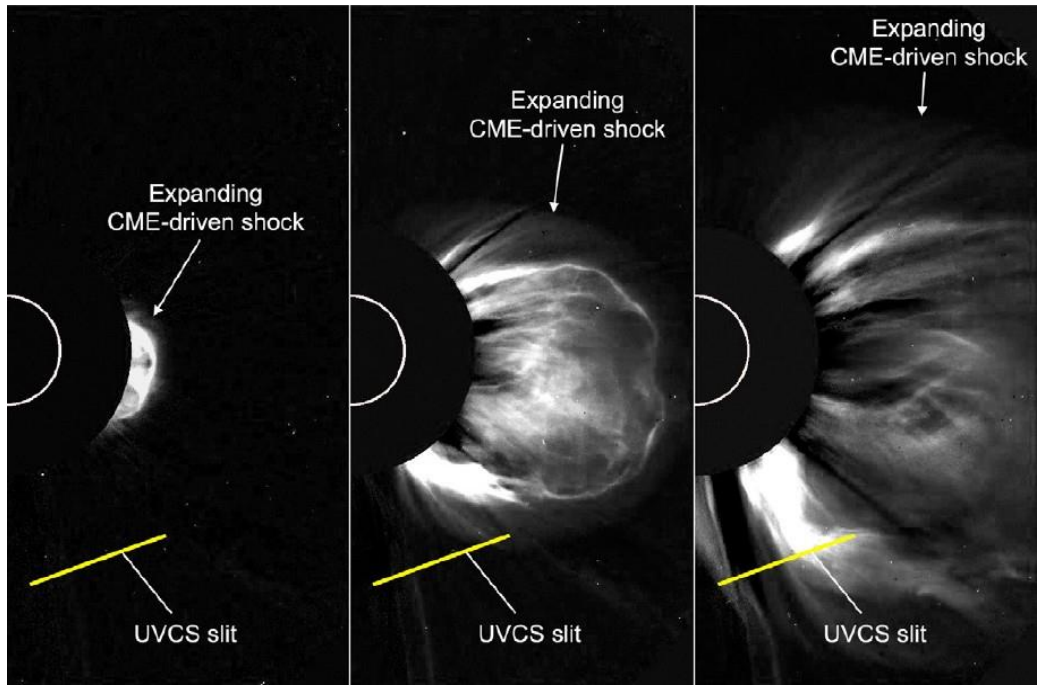

Figure 1: the expansion of a CME-driven shock wave as observed in the LASCO/C2 WL coronagraph (from Bemporad & Mancuso 2010).

## 1.2. The shock standoff distance technique

Gopalswamy & Yashiro (2011) first applied to remote sensing observations of CMEs this technique based on the measurement of the so called shock "standoff distance" $\Delta$, i.e. the distance between the shock driver (i.e. the expanding flux-rope) and the shock wave. Given $\Delta$ and the radius of curvature $R$ of the flux-rope, a semi-empirical formula relates the $\Delta/R$ ratio with the value of the shock sonic Mach number $M_s$. The $M_s$ value derived from the observed $\Delta/R$ ratio can be combined with the measured shock speed $v_{sh}$ to provide a measurement of the upstream Alfvén speed $v_A$, by assuming that the Alfvéninc Mach number $M_A \sim M_s$, and by assuming a value for the pre-shock solar wind speed $v_{out}$. Given also the measured pre-shock coronal electron density $n_e$, the magnetic field strength is provided by $B = v_A\,(\mu\,m_p\,n_e)^{1/2}$. Various authors applied this method to shocks observed with WL coronagraphs (Kim et al. 2012), EUV full disk imagers (Gopalswamy et al. 2012) and heliospheric imagers (Poomvises et al. 2012), providing measurements of the coronal field up to heliocentric distances larger than 120 solar radii.

Nevertheless, this technique has various limitations. As pointed out by Savani et al. (2012) there are some uncertainties in the expression of the semi-empirical relationship between the $\Delta/R$ ratio with $M_s$ which has two expressions:

$$\frac{\Delta}{R} = k\,\frac{(\gamma-1)M_s^2+2}{(\gamma+1)M_s^2} \quad \text{and} \quad \frac{\Delta}{R} = k\,\frac{(\gamma-1)M_s^2+2}{(\gamma+1)\left(M_s^2-1\right)}\;,$$

where the first expression was originally derived by Priest (1984) for the hydrodynamic (HD) shock case, while the second one is an adjustment made on intuitive basis by Farris & Russell (1994) to avoid the fact that the first expression is not valid under low Mach number regime ($M_s$ < 3). A second problem is that (as discussed in Savani et al. 2012) the value of the constant $k$ can vary depending on the oblateness of the shock driver (hence of the CME flux rope), which is usually a poorly constrained parameter (because of projection effects and unknown 3D shape of the CME) and is also expected to change as function of time/distance (Savani et al. 2011). The third problem is that (even though more appropriate MHD formulas were provided by Priest & Forbes 2007) the above expression holds for the HD case, and can be applied in the magneto-hydrodynamic (MHD) case if one assumes that $M_A \sim M_s$, hence $v_A \sim v_s$, while usually in the solar corona $v_A \gg v_s$. The HD solution is a good approximation for the MHD case when the magnetic pressure is much smaller than the plasma pressure ($\beta \gg 1$), but this is not the case for the lower corona and is true only for the interplanetary medium, hence for the analysis of heliospheric images or in situ measurements (Fairfield et al. 2001). Moreover, the application of the above method to CMEs observed with WL coronagraphs requires the assumption of another free parameter, the solar wind speed, to convert the measured shock speed into an upstream speed. Also, this technique is able to provide only a 1D radial profile of the magnetic field strength in the direction of propagation of the shock.

### 1.3. The WL-UV technique

To overcome these limitations, an alternative technique to derive coronal magnetic fields from shocks has been developed by the Group in Turin. In particular, Bemporad & Mancuso (2010) first demonstrated that WL and UV observations of a coronal shock (Figure 1) can be combined to infer the magnetic field strength and deflection across the shock front. In the technique pre- and post-shock densities are derived as usual from WL, while pre-shock solar wind velocities and plasma temperatures are derived from UV with standard techniques. The WL images are also used to estimate the shock speed and the inclination angle $\vartheta_{Bn}$ between the normal to the shock surface (projected on the plane of the sky) and the pre-shock magnetic field, assuming that above 2 solar radii it's radial. Given the shock compression ratio, shock inclination and velocity derived from WL, and the pre-shock plasma temperature and outflow velocity derived from UV, the MHD Rankine-Hugoniot equations written for the general case of an oblique shock provide not only the post-shock plasma temperatures, but also the post-shock outflow velocities, pre- and post-shock magnetic field strengths, as well as the velocity and magnetic field vector deflections across the shock. This technique, being a combination of WL and UV data, is very promising for application on future observations by the Metis coronagraph (Antonucci et al. 2012; Fineschi et al. 2012) on-board the forthcoming ESA-Solar Orbiter mission.

In a following work (Bemporad & Mancuso 2013) we introduced an empirical relationship to derive the Alfvénic Mach number $M_{A\angle}$ for the general case of an oblique shock directly from WL observations of shocks as

$$M_{A\angle} = \sqrt{\left(M_{A\perp} \sin \vartheta_{Bn}\right)^2 + \left(M_{A//} \cos \vartheta_{Bn}\right)^2}$$

where $M_{A\perp}$ and $M_{A//}$ are the Alfvénic Mach numbers for the special cases of perpendicular and parallel shocks, respectively, that are easily estimated from the shock compression ratio by assuming low plasma β condition. This empirical relationship, tested observationally (Bemporad, Susino & Lapenta 2014) and numerically (Bacchini et al. 2015), allowed the authors to derive for the first time the magnetic field strength in the corona crossed by a shock over a huge region covering 10 solar radii in altitudes and 110° in latitudes (Susino, Bemporad & Mancuso 2015).

This technique has also some uncertainties. First, it assumes that the pre-shock magnetic field is radial, an hypothesis which is more reliable above coronal streamer regions and usually not acceptable above polar regions, where super-radial expansion of the solar wind occurs. Second, it assumes that all the plasma parameters are derived on the plane of the sky, and (as it happens usually for remote sensing observations) it is hard to estimate the effect of the line of sight (LOS) integration in the determination of the shock compression ratio and all the other parameters derived from this quantity. Issues related with LOS integration are present in all kinds of remote sensing observations of optically thin plasmas. Nevertheless, we point out that in the above determinations of the shock compression ratios from WL the authors showed how to take into account different LOS extensions of the shocked plasmas at different altitudes (see e.g. Bemporad & Mancuso 2010), thus improving the simplistic usual assumption by previous authors of a constant LOS thickness (e.g. Ontiveros & Vourlidas 2009).

## 2. Present debates on magnetic field measurements with shock

After a first excitement by part of the community for the possibility to directly observe the propagation of shocks in the corona, the validity of these results are becoming more debated. It has been pointed out that the observed hemispherical WL features expanding ahead of CMEs could not be unambiguously differentiated from the front of compression waves, due to the pileup of coronal plasma lying above the expanding CME. The hypothesis that these features are really shocks is better supported when at the same time at least one of these observational features is also reported:

- remote sensing observation in radio data of a type-II burst;
- remote sensing observation on EUV-UV spectroscopic data of non-thermal line broadening ahead of the CME front;
- in situ observation of a gradual SEP event.

Another observational property supporting the real formation of a CME-driven shock is provided directly by the speed of the CME front as observed in WL images. In fact, the formation of a shock is expected only when the driver (i.e. the expanding CME) is moving in the corona faster than the local Alfvén, sound and/or magneto-sonic speeds. In agreement with this, faster CMEs are also statistically more associated with type-II radio bursts (Gopalswamy et al. 2010) and a good correlation exists between CME speeds and SEP fluxes (Kahler & Vourlidas 2013). In any case, the detection of one or more of these features is not telling us where the shock is formed in the corona, and the possibility that the observed WL feature is not the real shock cannot be completely ruled out.

A criticism often also pointed out is that, if these hemispherical features are really the shocks, then the intensity variation plotted perpendicular to the shock should contain a steep discontinuity, similar to those observed in density measurements acquired by in situ spacecraft at shock transit. On the other hand, radial plots of the WL intensity across these hemispherical features can show a gradual increase of the WL, hence a gradual increase in the electron column density of the plasma. Nevertheless, this is only an apparent inconsistency due to a combination of the 3D geometry of the shock surface and the integration along the LOS through the optically thin coronal plasma. In fact, the shock surface will have an almost hemispherical shape (as demonstrated by 3D numerical simulations), and the location of the shock boundary in the 2D projected view will correspond to the pixels where the lines of sight graze this surface. Then, moving radially into the shock sheath, the fraction of the LOS intercepting the sheath region will progressively increase, leading to the observed progressive increase in the WL intensity, as nicely shown by Manchester et al. (2008).

More in general, the above techniques require many assumptions to derive the magnetic fields from the observed WL images, making it difficult to estimate the real accuracy of these measurements. We suggest that these uncertainties will be better constrained by combining data analysis with forward modelling of synthetic observations (e.g. Gibson 2015).

## 3. Conclusions and future perspectives

Despite the uncertainties discussed here, both the "standoff" and "WL-UV" techniques (recently applied for inter-comparison to the same event by Susino et al. 2015) are now providing reasonable measurements of the coronal magnetic fields over broad intervals in altitudes and latitudes never reached before. Comparison with other observational methods and the analysis of synthetic data will likely help us to improve and optimize these techniques in the future. A limit of these techniques is that field measurements are provided only after the eruption responsible for the shock wave: hence these methods give an "a posteriori" knowledge of the pre-CME coronal fields, and likely will be applicable for forecasting purposes only when statistical analyses will be carried out.

**Acknowledgments**

R. Susino acknowledges support from ASI contract I/013/12/0-1; F. Frassati acknowledges support from INAF Ph.D. Grant.

**References**

Antonucci, E.; Fineschi, S.; Naletto, G.; Romoli, M.; Spadaro, D.; Nicolini, G.; et al. "Multi Element Telescope for Imaging and Spectroscopy (METIS) coronagraph for the Solar Orbiter mission", 2012, Proc. SPIE, 8443, 844309, 12 pp.

Bacchini, F.; Susino, R.; Bemporad, A.; Lapenta, G.; et al. "Plasma Physical Parameters along CME-driven Shocks. II. Observation-Simulation Comparison", 2015, ApJ, 809, 58-70.

Bemporad, A., & Mancuso, S., "First Complete Determination of Plasma Physical Parameters Across a Coronal Mass Ejection-driven Shock", 2010, ApJ, 720, 130-143.

Bemporad, A., & Mancuso, S., "Super- and sub-critical regions in shocks driven by radio-loud and radio-quiet CMEs", 2013, JAdR, 4, 287-291.

Bemporad, A., Susino, R., & Lapenta, G., "Plasma Physical Parameters along Coronal-mass-ejection-driven Shocks. I. Ultraviolet and White-light Observations", 2014, ApJ, 784, 102-112.

Chifu, I.; Inhester, B.; Wiegelmann, T., "Coronal magnetic field modeling using stereoscopy constraints", 2015, A&A, 577, id.A123, 8 pp.

Fairfield, D. H.; Cairns, Iver H.; Desch, M. D.; Szabo, A.; Lazarus, A. J.; Aellig, M. R.; "The location of low Mach number bow shocks at Earth", 2001, 106 (A11), 25361-25376.

Farris, M. H., and Russell, C. T. (1994). Determining the standoff distance of the bow shock: Mach number dependence and use of models. J. Geophys. Res. 99, 17681-17689. doi: 10.1029/94JA01020.

Fineschi, Silvano; Antonucci, Ester; Naletto, Giampiero; Romoli, Marco; Spadaro, Daniele; Nicolini, Gianalfredo; "METIS: a novel coronagraph design for the Solar Orbiter mission", et al. 2012, Proc. SPIE, 8443, 84433H, 13 pp.

Gibson, S., "Data-model comparison using FORWARD and CoMP", 2015, Proc. IAU, 305, 245-250.

Gopalswamy, N.; Xie, H.; Mäkelä, P.; Akiyama, S.; Yashiro, S.; Kaiser, M. L.; et al., "Interplanetary Shocks Lacking Type II Radio Bursts", 2010, ApJ, 710, 1111-1126.

Gopalswamy, N., & Yashiro, S., "The Strength and Radial Profile of the Coronal Magnetic Field from the Standoff Distance of a Coronal Mass Ejection-driven Shock", 2011, ApJ, 736, L17, 5 pp.

Gopalswamy, Nat; Nitta, Nariaki; Akiyama, Sachiko; Mäkelä, Pertti; Yashiro, Seiji, "Coronal Magnetic Field Measurement from EUV Images Made by the Solar Dynamics Observatory", 2012, ApJ, 744, 72, 7 pp.

Kahler, S.W., & Vourlidas, A., "A Comparison of the Intensities and Energies of Gradual Solar Energetic Particle Events with the Dynamical Properties of Associated Coronal Mass Ejections", 2013, ApJ, 769, 143, 12 pp.

Kim, R.-S.; Gopalswamy, N.; Moon, Y.-J.; Cho, K.-S.; Yashiro, S., "Magnetic Field Strength in the Upper Solar Corona Using White-light Shock Structures Surrounding Coronal Mass Ejections", 2012, 746, 118, 8 pp.

Kozarev, K. A.; Korreck, K. E.; Lobzin, V. V.; Weber, M. A.; Schwadron, N. A., "Off-limb Solar Coronal Wavefronts from SDO/AIA Extreme-ultraviolet Observations—Implications for Particle Production", 2011, ApJ, 733L, 25, 7 pp.

Manchester, Ward B., IV; Vourlidas, Angelos; Tóth, Gábor; Lugaz, Noé; Roussev, Ilia I.; Sokolov, Igor V.; et al., "Three-dimensional MHD Simulation of the 2003 October 28 Coronal Mass Ejection: Comparison with LASCO Coronagraph Observations", 2008, ApJ, 684, 1448-1460.

Ontiveros, V., & Vourlidas, A., "Quantitative Measurements of Coronal Mass Ejection-Driven Shocks from LASCO Observations", 2009, ApJ, 693, 267-275.

Priest, E. (1984). Solar Magneto-hydrodynamics. Geophysics and Astrophysics Monographs, Dordrecht: Reidel.

Poomvises, Watanachak; Gopalswamy, Nat; Yashiro, Seiji; Kwon, Ryun-Young; Olmedo, Oscar, "Determination of the Heliospheric Radial Magnetic Field from the Standoff Distance of a CME-driven Shock Observed by the STEREO Spacecraft", 2012, ApJ, 758, 118, 6 pp.

Priest, E., & Forbes, T., "Magnetic Reconnection: MHD Theory and Application", 2007, Cambridge University Press, 612 pp.

Savani, N. P.; Owens, M. J.; Rouillard, A. P.; Forsyth, R. J.; Kusano, K.; Shiota, D.; Kataoka, R., "Evolution of Coronal Mass Ejection Morphology with Increasing Heliocentric Distance. I. Geometrical Analysis", 2011, ApJ, 731, 109, 6 pp.

Savani, N. P.; Shiota, D.; Kusano, K.; Vourlidas, A.; Lugaz, N., "A Study of the Heliocentric Dependence of Shock Standoff Distance and Geometry using 2.5D Magnetohydrodynamic Simulations of Coronal Mass Ejection Driven Shocks", 2012, ApJ, 759, 103, 11 pp.

Susino, R., Bemporad, A., & Mancuso, S., "Physical Conditions of Coronal Plasma at the Transit of a Shock Driven by a Coronal Mass Ejection", 2015, ApJ, 812, 119, 13 pp.

Vourlidas, A.; Wu, S. T.; Wang, A. H.; Subramanian, P.; Howard, R. A., "Direct Detection of a Coronal Mass Ejection-Associated Shock in Large Angle and Spectrometric Coronagraph Experiment White-Light Images", 2003, ApJ, 598, 1392-1402.

Vourlidas, A.; Lynch, B. J.; Howard, R. A.; Li, Y., "How Many CMEs Have Flux Ropes? Deciphering the Signatures of Shocks, Flux Ropes, and Prominences in Coronagraph Observations of CMEs", 2013, Sol.Phys., 284, 179-201.